\documentclass[]{spie}  

\usepackage{amsmath,amsfonts,amssymb}
\usepackage{graphicx}
\usepackage[colorlinks=true, allcolors=blue]{hyperref}

\title{A user-friendly tool to convert photon counting data to the open-source Photon-HDF5 file format}

\author[a,b]{Donald Ferschweiler}
\author[c]{Maya Segal}
\author[c]{Shimon Weiss}
\author[c]{Xavier Michalet}

\affil[a]{Mathematics, Sciences, and Engineering Division, Rio Hondo College, Whittier, CA, USA}
\affil[b]{Department of Physics and Astronomy, California State University, Long Beach, CA, USA}
\affil[c]{Department of Chemistry and Biochemistry, University of California, Los Angeles, CA, USA}

\authorinfo{Send correspondence to X.M. E-mail: michalet@chem.ucla.edu}

\begin{document} 
\maketitle

\begin{abstract}
  Photon-HDF5 is an open-source and open file format for storing photon-counting
  data from single molecule microscopy experiments, introduced to simplify data
  exchange and increase the reproducibility of data analysis. Part
  of the Photon-HDF5 ecosystem, is \texttt{phconvert}, an extensible python
  library that allows converting proprietary formats into Photon-HDF5
  files. However, its use requires some proficiency with
  command line instructions, the python programming language, and the YAML
  markup format. This creates a significant barrier for potential users without
  that expertise, but who want to benefit from the advantages of releasing their
  files in an open format. In this work, we present a GUI that lowers this
  barrier, thus simplifying the use of Photon-HDF5. This tool uses the
  \texttt{phconvert} python library to convert data files originally saved in
  proprietary data formats to Photon-HDF5 files, without users having to write a
  single line of code. Because reproducible analyses depend on essential
  experimental information, such as laser power or sample description, the GUI
  also includes (currently limited) functionality to associate valid metadata
  with the converted file, without having to write any YAML. Finally, the GUI
  includes several productivity-enhancing features such as whole-directory batch
  conversion and the ability to re-run a failed batch, only converting the files
  that could not be converted in the previous run.
\end{abstract}

\keywords{photon-counting, single-molecule, fluorescence, open-source,
Photon-HDF5, phconvert, GUI, Python}

\section{INTRODUCTION}
\label{sec:intro}  

\subsection{Photon-HDF5 File Format}

The single-molecule fluorescence spectroscopy (SMFS) community has recently promoted
some level of standardization in measurement practices and analysis pipelines \cite{EL2021}.
In order to enable reproducibility of data analysis and thus strengthen trust in
scientific results, it is important for data to be freely and easily available, for example by depositing it in online repositories.
Moreover, data should be readable and well-documented, which is best done using standardized
file formats and abundant, human-readable metadata.

Photon-counting SMFS methods implement widely varying setup geometries and
hardware. Common photon-counting techniques include single-molecule Förster
Resonance Energy Transfer (smFRET), fluorescence correlation and
cross-correlation spectroscopy (FCS and FCCS), and fluorescence lifetime
spectroscopy. In the case of smFRET, setups may have a variety of different
features, for example, multi-color detection schemes, nano or microsecond
alternative laser excitation (ns- or $\mu$s-ALEX), or polarization anisotropy. In
each of these cases, parameters change between measurements such as the number
and wavelength of light sources, the laser period in setups using alternation,
number of detectors, or the presence of polarizers, etc.

In addition to the wide variety of hardware, there is a wide variety of acquisition software and file formats. 
Data is often recorded using custom-built binary file formats, either supplied by an equipment
manufacturer or designed in-house by the laboratory in which the experiment was
performed. Oftentimes, the former are preferable to the latter, as manufacturers
have a vetted interest in long-term support of their customers, while graduate
students and postdocs follow each other in a lab, with variable degree of overlap. 
However, manufacturers understandably
develop their own formats adapted to their hardware and market niche, with little
concern for compatibility with other products.
The result is a slew of file formats, some well-documented, some not at all, which
leads to loss of effective access to the data, even when the files remain available.
Unifying common data storage formats helps to alleviate the issues associated
with proprietary file formats, and as the number and variety of techniques
continues to grow, it becomes increasingly important to standardize an open and
accessible format for the burgeoning SMFS community.

A few years ago, we introduced a open and standardized photon-counting file format, Photon-HDF5, based
on the HDF5 open source file format~\cite{BJ2016}. The HDF5 
organization is a vibrant community supporting multiple operating systems, and provides
a free HDFView software for these platforms, enabling users to open, read, and copy
data from these files without knowing any programming, making it an ideal
vehicle for universal data storage~\cite{hdf5_group,hdfview}.
Based on these foundations, the
Photon-HDF5 file format formalizes the organization of SMFS data obtained with 
photon-counting hardware, by logically storing raw data along with metadata for the sample and
measurement. Storing data and metadata together is extremely important for long-term reproducibility, 
see Photon-HDF5.org or~\href{http://photon-hdf5.github.io/}{http://photon-hdf5.github.io/} for a complete documentation.

By virtue of the HDF5 format's built-in features, Photon-HDF5 incorporates
efficient data compression, and can be easily read within computing environment
programming languages such as Python or MATLAB. Furthermore, accompanying the
format's detailed description, code snippets for MATLAB and LabVIEW are provided
as examples, as well as a python based file converter called \texttt{phconvert}
(\href{http://photon-hdf5.github.io/phconvert/}{http://photon-hdf5.github.io/phconvert/})~\cite{phconvert}.
The format is supported by an increasing number of SMFS data processing
packages, such as \texttt{FRETBursts} (python)~\cite{PO2016} and \texttt{ALiX}
(LabVIEW)~\cite{PO2017}.

\subsection{phconvert}

\texttt{phconvert} is a helper library which was released together with
Photon-HDF5, designed to convert some common proprietary file formats to the
Photon-HDF5 format. Currently supported raw data files include PicoQuant's
.HT3/.PT3/.PTU/.T3R, Becker \& Hickl's .SPC/.SET, and the .SM format used by our
lab and others for $\mu$s-ALEX smFRET. As mentioned earlier, one advantage of
such a conversion is that files can be viewed in human readable forms using
HDFView~\cite{hdfview}, and most importantly, when done properly, the data is
stored with relevant metadata, helping to make sense of the experimental and
measurement parameters.

Using \texttt{phconvert} requires basic familiarity with Python, something which
is becoming increasingly common in the community, but remains a potential obstacle to the
adoption of this simple action: converting ones' data files into Photon-HDF5 formatted files for deposition into a public online repository. Additionally, some understanding of the different metadata fields of Photon-HDF5 (and more generally, an understanding of the general logic of the format organization)
is required to generate fully documented files. Finally, the metadata to be provided to \texttt{phconvert} needs to be prepared in a YAML file, which requires some knowledge of this simple yet specific data serialization language (YAML Ain't Markup Language™)~\cite{yaml}.

In order to lower the barrier to adoption of Photon-HDF5 for sharing
photon-counting data generated in SMFS experiments, we have developed a
graphical user interface (GUI) for the \texttt{phconvert} library that
significantly reduces the need for the specialized technical skills currently
required for file conversion. In this paper, we will describe the current state
of the GUI, detailing some of the ways it can currently be used, its
limitations, and future improvements that could be made.

\section{Software design}

\subsection{Features}

The goals set for this tool were to:

\begin{description}
  \item[Simplify installation by packaging all dependencies]
        To obtain \texttt{phconvert}, users have several options, all of which
        involve technical knowledge. They can download the package from GitHub,
        taking care to properly install it so that it plays well with the system's
        Python installation, or they can use a package manager such as
        \texttt{pip} or \texttt{conda}. We wanted to improve on this experience
        for users without technical experience. Most users have likely not heard
        of these tools, let alone feel comfortable using them. So installation
        of the GUI should be as straight-forward as possible, and not require
        mucking about with language package managers or stuff like the \texttt{PYTHON\_PATH} variable.
  \item[Provide a user-friendly conversion tool] Minimal prior knowledge should
        be presumed of the user. Especially no technical knowledge regarding
        programming, the command line, YAML, etc.  Common operations should be obvious,
        so that the user is not left hunting for what they want. The correct thing to do
        should be made as easy as possible, and there should be no hidden suprises that
        sabotage users.
  \item[Support batch conversion of multiple files]
        Users may have more than one file that they wish to convert, and it can
        be convenient to deal with a large directory all at once. To streamline
        the process of converting many files at a time, there should be batch
        conversion mode.
  \item[Provide helpful feedback during and after conversion]
        As a minimum requirement, there should be a way for the users to know whether the conversion process
        was successful. Users should also be able to
        find out what when wrong in the case of a failure, so that they can fix
        the situation and try again.
\end{description}

\subsubsection{Additional nice-to-have features} 
\begin{description}
  \item[YAML-free metadata specification]
        Many users won't know what YAML is, nor the syntax for it, or even how
        to create a YAML file in the first place. A method for users to provide
        the metadata for the data file(s) being converted to Photon-HDF5 without
        having to use YAML directly would definitely be useful.
  \item[Support for multiple platforms]
        For widest adoption, the GUI should be available on all the common
        platforms so that most researchers who want to use it can.
  \item[Seamless upgrades to future \texttt{phconvert} versions]
        The user should be insulated from worrying about differences in the
        underlying library version, and the GUI should be developed in such a
        way that the maintainer(s) have as smooth a time as possible when the
        time comes to update.

\end{description}

\subsection{Implementation}

To make things truly point-and-click, a thin wrapper around
phconvert was needed that provided a single entry point
for conversion. To make the interface as simple and discoverable 
as possible, a minimum amount
of graphical elements were added, and a focus was placed on trying to make the things that
should be easy, as easy as possible. The button to convert a file or batch is the biggest, most
visible button. There is a status bar for each conversion task that lets the
user know if conversion succeeded, and the status bar for batches tries to be
informative regarding both the current attempt and the overall status of the
batch.

Implementing the resume feature
for batch conversions required keeping track of the files that had been
converted already, and to not abort the whole conversion process when an error is encountered. To
deal with both of these, a basic logging system was implemented. When a
directory is selected for batch conversion, a log file is created inside it. As the
software goes through the directory trying to convert (sensible) files, it
writes to the log whether each file was successful or not, and the reason for
failure. If not all files are successful, this log can be used by the GUI to
reconstruct the list of files it already converted successfully, saving time for
the user by only converting the files in the batch that failed, and which were
presumably fixed. It is also intended to be useful to users in finding out why a
conversion failed, and logs printed representations of all errors encountered.

\subsection{Tools used}

\subsubsection{GUI design}

The GUI is built using Python's Tkinter toolkit~\cite{tkinter}. The primary reason this
was chosen was because of all the graphical user interface toolkits available to
Python, Tkinter appeared to support the most interactivity during development. A high
degree of interactivity can improve development productivity by reducing the
time between writing code and seeing the results, and in the ideal case allows
updates on the fly so that bugs can be fixed literally seconds after identification.
Interactivity becomes even more valuable when there is significant
uncertainty in the development process, such as when developing a novel
graphical user interface for a workflow previously based on the command line and
Python programming. To this end, a prototype of the GUI was
developed with the Common Lisp language and
the Common Lisp Omnificent GUI (CLOG) toolkit~\cite{clog}. 
when the code was changing quickly, CLOG was a fast and convenient way to develop the GUI. 
After the initial development steps, the GUI was ported to the target language and toolkit,
Python and Tkinter. 

\subsubsection{GUI distribution}
Once the GUI was ready to be distributed to others for testing, a new set of
tools for deployment were required. Since we are writing this GUI for users who
are not expected to manually install a Python package and dependencies with a
package manager like \texttt{pip} or \texttt{conda}, the GUI should be
distributed as a self-contained executable that the user need only download and
double click to get started. To achieve this, we used \texttt{pyinstaller} to
package up the GUI code along with all of its dependencies, including the Python
interpreter, with the exception of things such as \texttt{glibc} which are
provided by the operating system. At this early stage, the GUI is packaged as an
executable within a zipped up directory structure that contains all the
mentioned dependencies. This permits easier debugging. In later stages, a single
self-contained executable, rather than the executable plus dependencies folder
used for development, will be created and released on Github, all that is needed
being a command flag in the build process.

\subsubsection{Build automation}

In order to provide multi-platform releases (at least on the most popular ones
in the SMSF community, namely Linux, Mac, and Windows) without the burden 
of supporting dedicated machines  for the builds, the AppVeyor online platform was used~\cite{appveyor}.
As part of getting software into the hands of testers, a continuous integration
(CI) pipeline connected to the GUI's Github repository was developed using AppVeyor,
starting from the configuration used by \texttt{phconvert}. Each time a commit is pushed,
the pipeline builds the executable bundles on both Linux and Windows, then
uploads them to the Github releases page. Initially, we intended to release an executable
running on macOS, but unresolved issues related to Apple's security
policies have led to us to postpone it. Contributions towards resolving these
problems are welcome.

While developing the CI pipeline, the requirement that the code snippets in the
configuration file be runnable on every platform being targeted, except if
flagged otherwise, initially caused some issues. The crux of the matter was that
each platform had a different shell. Windows had
\texttt{batch}/\texttt{cmd.com}, Linux had \texttt{Bash}, and Mac had
\texttt{Zsh}. At the face of it, this meant that large chunks of code would need
to be copied between the different platforms' build processes, each slightly
tweaked for the specific platform. This is bad from a maintainability
standpoint, since it multiplies the amount of work necessary to make a single
change and the repetitiveness of the code provides a place for bugs to hide and
the opportunity for things to be forgotten to be changed. What was wanted was a
way to write the build process code in the AppVeyor configuration file such that it
would be valid on all platforms, with as much reuse as possible so that only the
things which absolutely needed to be different (such as installation of software
on Linux and Mac that was already present on Windows) were different. Luckily,
one common denominator across all the AppVeyor images used is the installation
of the \texttt{Powershell} or \texttt{Powershell Core} (\texttt{pwsh}) shell.
Additionally, AppVeyor provides a means in the configuration file to mark code as
\texttt{pwsh} snippets/blocks to be run. This provided the necessary common
language between platforms. As of now, all code in the AppVeyor configuration file with
the exception of one special case is written in \texttt{pwsh}, for better
maintainability.

\section{Software demonstration}


In the following subsections, we will demonstrate the GUI's basic features by
illustrating how files can be converted to Photon-HDF5.

\subsection{Installation}

The current installation process for the GUI is as follows:
\begin{enumerate}
  \item Go to
        \href{https://github.com/ramenbytes/phui}{https://github.com/ramenbytes/phui},
        which is the working repository for the GUI right now. PHUI stands for
        \textbf{Ph}convert \textbf{U}ser \textbf{I}nterface.
  \item Click on \textbf{Releases}.
  \item Download the zip file of your choice corresponding to your platform and
        chosen Python version. For Linux, choose an Ubuntu-labeled release, and
        for Windows, choose a Visual Studio one.
  \item Extract the downloaded zip file.
  \item Inside the zip folder, descend into the \texttt{dist/gui} folder.
  \item Double-click the \texttt{gui} executable to run the GUI.
\end{enumerate}

The only system requirements are that the user be on a platform at least as
up-to-date as the Appveyor image used to build the executable. Currently, that
is Ubuntu 18.04.4 LTS (Bionic Beaver) and Windows 10. They do not need
to be on the \emph{same} system, just one whose system libraries (such as
\texttt{glibc} on Linux) are contemporary to the build image's. The rest of the
dependencies are taken care of by the build process and are stashed in the
distributed folder containing the executable which the user downloads.

\subsection{Single-File Conversion}
\label{sec:single-file}

\begin{figure}[ht] \centering
\includegraphics[width=\textwidth]{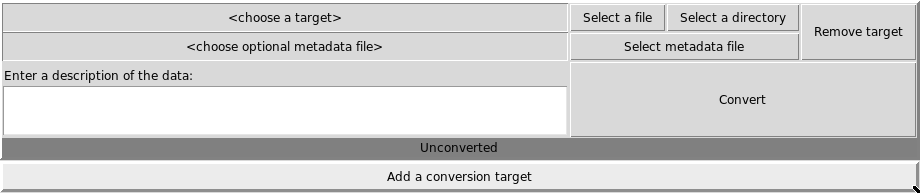}
  \caption{Initial window shown to users upon starting the GUI.}
  \label{fig:init-window}
\end{figure}

The simplest use case for the GUI is the conversion of a single file. Upon
starting the GUI, the user is greeted by the window in Figure
\ref{fig:init-window}. Starting in the top left, the initially empty
\textbf{choose a target} field displays the filename of the data file the user
wants to convert. To select a file for conversion, the user clicks on the
\textbf{Select a file} button to the immediate right of the field. The user is
presented with a file dialog which they can use to navigate to and select the
desired file. After a file is chosen, metadata needs to be associated with it.
There are currently several methods of doing so, and more planned.
\begin{itemize}
  \item First, the user can provide a description in the white textbox near the
        bottom of the window. The Photon-HDF5 format requires a description, so
        the conversion process will fail if one isn't provided.
  \item Second, \texttt{phconvert} provides default values for certain metadata
        fields. Absent user input, these values will be used. To override these
        default values or to provide values for other fields, users currently
        need to use the third method,
  \item Third, clicking the \texttt\\bf{Select metadata file} button to choose a
        YAML file containing the relevant info. This is intended as a stopgap
        solution until a graphical metadata editor is developed, freeing the
        users from worrying about YAML files. However, in its current state, the
        conversion process is not entirely YAML-free.
    
\end{itemize}

\begin{figure}[ht] \centering
\includegraphics[width=\textwidth]{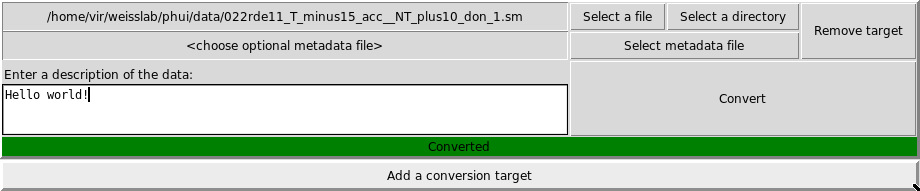}
  \caption{A successful conversion}
  \label{fig:successful-single-conversion}
\end{figure}

After a data file has been chosen, and metadata associated with it, users will
need to press the \textbf{Convert} button. If the conversion is successful (no
errors are thrown during the process), then the status bar near the bottom of
the window (grey, with the label \textbf{Unconverted} in Fig.
\ref{fig:init-window}) will change to green, as shown in Fig.
\ref{fig:successful-single-conversion}.

\begin{figure}[ht] \centering
\includegraphics[width=\textwidth]{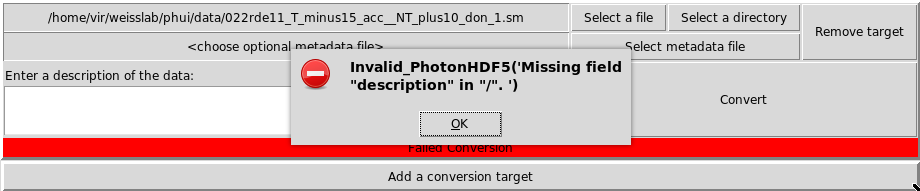}
  \caption{An unsuccessful conversion}
  \label{fig:failed-single-conversion}
\end{figure}

If an error occurs during the conversion process, the status bar changes to red
with a message noting the failure, and a dialog window with a printed
description of the error is presented to the user as seen in 
Fig.~\ref{fig:failed-single-conversion}.

The conversion can be retried by closing the error dialog, and pressing
\textbf{Convert} again. If the user fixes the error before trying again, the
conversion should succeed this time around.

\subsection{Batch Conversion}

Directly to the right of the \textbf{Select a file} button is the \textbf{Select
a directory} button, as seen in Fig. \ref{fig:init-window}. This button allows a
user to select an entire directory as a candidate for conversion, as opposed to
a single file. When the user attempts to convert the directory, a subroutine
will gather a list of all the files in the provided directory that can be
converted, based on the filename extensions. Everything else is excluded from
the conversion attempt. Then, one by one, the program tries to convert each of
these files, recording in a log whether a file conversion was successful, and if
not, what the error was.

\begin{figure}[ht] \centering
\includegraphics[width=\textwidth]{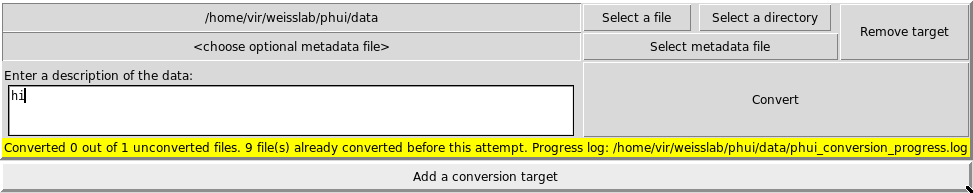}
  \caption{Status from a partial batch conversion. At the start of this attempt,
one file was unconverted. Nine were logged as previously converted. This attempt
failed to convert the single unconverted file. The progress log at the provided
path lists the error which resulted in the failure.}
  \label{fig:partial-batch-status}
\end{figure}

A user can try re-running the batch by clicking \textbf{Convert} again; this
time, only the files marked as failures in the log will be tried again. This is
designed such that long batch conversions are not repeated needlessly. At the
end of each attempt, the status bar shows the fraction of unconverted files that
were successfully converted on this attempt, how many were converted in previous
attempts, and where to find the log file. See Fig.
\ref{fig:partial-batch-status} for an example.

The metadata for batch conversions is obtained in the same way as for single
file conversions, however it will be applied to \emph{all} files in the batch.
That means that currently, batch conversions are only suitable for groups of
files that have identical metadata. For example, data from several runs of
the 
same experiment. Offering more flexibility in different scenarios is a top
priority for future development of the GUI.

\subsection{Multiple Targets}

\begin{figure}[ht] \centering
\includegraphics[width=\textwidth]{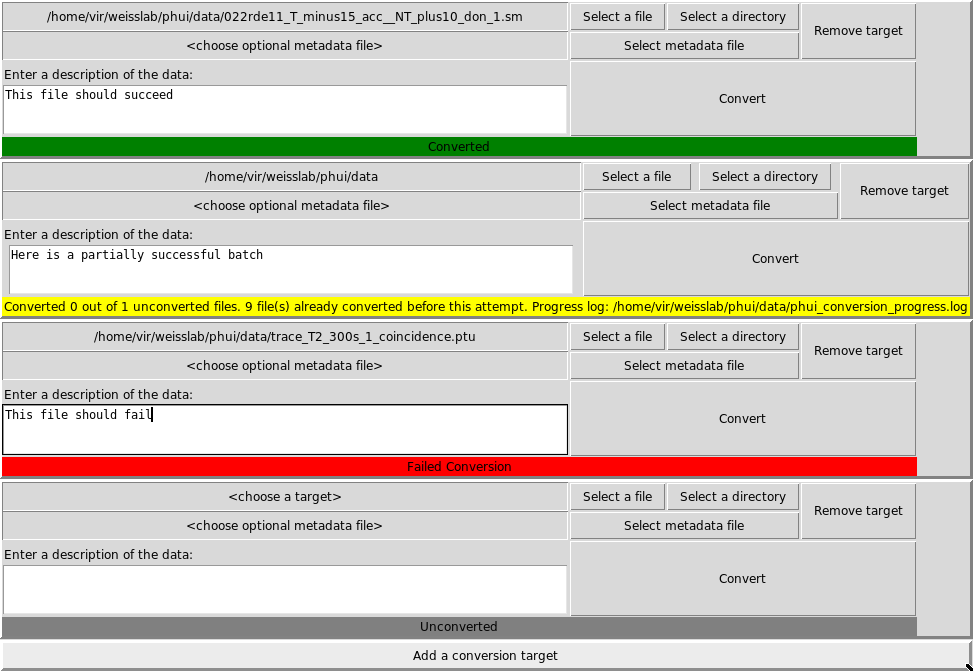}
  \caption{Multiple targets for conversion. One successful file (top), one
    partially successful batch (second from top), one failed file (second from
    bottom), and one target that hasn't been attempted yet and does not have any data or
    metadata (bottom).}
  \label{fig:multiple-targets}
\end{figure}

The last two buttons of interest are \textbf{Remove target} in the top right, and
\textbf{Add a conversion target} on the bottom of Fig.~\ref{fig:init-window}. These
allow the addition of additional individual conversion ``tasks'' for the GUI to
perform. Adding targets allows the user to add files or batches of
files, each of which is handled individually with its own associated metadata,
status bar, and button for conversion. See Fig.~\ref{fig:multiple-targets} for
an example.

Using the \textbf{Remove target} button, users can remove files or batches of files
from the GUI's display, for whatever the reason may be.

\section{CONCLUSION}

We have developed a GUI which lowers, but does not yet
eliminate, the barriers faced by users in the adoption of Photon-HDF5.
Absolutely no Python or command line experience is necessary to use the GUI, 
which is certainly an improvement. 
The primary remaining barriers are in metadata storage and discoverability.

\subsection{Metadata}

The GUI allows users to associate metadata with the data sets being
converted, but in a very limited manner. Anything different than the defaults
provided by \texttt{phconvert} must be provided via the fallback approach, i.e., manually
entering the metadata into a YAML file using a text editor. 
Unfortunately, because of the variety of
different experimental setups in the photon-counting community, most data sets
will not perfectly match the defaults provided and the YAML file option will
need to be used to correct the default values.

At this stage, it is also not
immediately apparent to the user that there are default values for fields, or
even which fields exist. As of now, nothing is displayed in the GUI that
indicates the presence of these fields and that the defaults from \texttt{phconvert} are being
used. This means that if a user is not familiar with the GUI, 
it is very easy for them to be fooled into thinking that
they have successfully converted a file to Photon-HDF5, when in fact that is
only partially true. While their experimental data may have successfully been
converted, it is now associated with the default values provided by \texttt{phconvert},
which are not guaranteed to make sense in the user's specific case.

However, this temporary limitation can easily be avoided since most users will typically
generate similar files with the hardware at their disposal. Therefore, proper use of the 
GUI will be guaranteed if the users use the same YAML file, adapted to their setup, simply
modifying the few fields identifying each experiment before running the GUI.

Since the GUI allows entering a new data description for each new converted file,
we believe that this functionality of the software will help a number of laboratories to
convert their files, once they are in possession of an appropriate template YAML file.
We are planning to provide a few examples of such YAML files with the first release, 
and encourage users to publicize theirs, together with a detailed description of their setups,
so that the community can benefit from best practices from the leading labs.

Future developments of the GUI will explore the possibility of implementing a
basic metadata editor in the GUI, which presents all the fields to the user and
allows them to enter values into them.

A related issue is that for batch conversions, there is currently
no way of specifying metadata on a per-file basis. This restricts the set of
data files which can be usefully and correctly batch converted to those which
are differing minimally or not at all. If even one metadata field is different between two
files, they will have to be processed in separate batches or individually. Ideally, users would be able to
have a fine-grained control over the metadata for each file in a batch, perhaps
being able to set defaults for the entire batch and overriding or supplementing
them on a per-file basis.

Finally, even if users can easily associate any metadata they want with a file
and/or batch, they still need to know what fields are required in the first place. 
The Photon-HDF5 specification is quite rich, and learning what metadata needs to be
associated with the data from which experimental setup is challenging for a first time user.
Beyond the suggestion made above of using a standardized YAML file for a given setup or set of experiments,
one possible (partial) solution would be to add a workflow in the GUI
that walks the user through the process, guiding them towards the fields they
need to include.

\subsection{Documentation}

A definite possibility for future improvements would be documentation for the
GUI, complete with clear instructions displayed in the repository
\texttt{README} file detailing how to install the GUI on each of the supported
platforms and the system requirements for each.

\subsection{Releases and Contributions}

This project is hosted on Github at
\href{https://github.com/ramenbytes/phui}{https://github.com/ramenbytes/phui},
and the executables and source code can be found there. During this stage of
development, all contributions such as pull requests or issues should be opened
against this repository. However, it is hoped that in the future \texttt{PHUI}
will be merged with the \texttt{phconvert} repository, in which case a notice
will be posted to the above repository directing users to the new location.

\section{ACKNOWLEDGEMENTS}

This work was funded in part by NIH grant R01 GM130942, NSF awards MCB 1818147
and EAGER 1842951, and by HFSP grant RGP0061/2019.

\noindent
DF is grateful for SPIE's and MKS Instruments' Student Conference Support
Program grant.

\noindent
DF also thanks Rio Hondo College, the UCLA-RHC Scientific Exchange, and
Christian Vaca for initiating this collaboration, and for providing support and
mentorship throughout.

\bibliography{manuscript} \bibliographystyle{spiebib}

\end{document}